\documentclass[final]{aipproc}
\usepackage{amssymb}

\layoutstyle{6x9}

\begin{document}

\title{Short and canonical GRBs}


\author{Carlo Luciano Bianco}{address={ICRANet, Piazzale della Repubblica 10, 65122 Pescara, Italy.}, altaddress={Dipartimento di Fisica, Universit\`a di Roma ``La Sapienza'', P.le Aldo Moro 5, 00185 Roma, Italy.}}

\author{Maria Grazia Bernardini}{address={ICRANet, Piazzale della Repubblica 10, 65122 Pescara, Italy.}, altaddress={Dipartimento di Fisica, Universit\`a di Roma ``La Sapienza'', P.le Aldo Moro 5, 00185 Roma, Italy.}}

\author{Letizia Caito}{address={ICRANet, Piazzale della Repubblica 10, 65122 Pescara, Italy.}, altaddress={Dipartimento di Fisica, Universit\`a di Roma ``La Sapienza'', P.le Aldo Moro 5, 00185 Roma, Italy.}}

\author{Maria Giovanna Dainotti}{address={ICRANet, Piazzale della Repubblica 10, 65122 Pescara, Italy.}, altaddress={Dipartimento di Fisica, Universit\`a di Roma ``La Sapienza'', P.le Aldo Moro 5, 00185 Roma, Italy.}}

\author{Roberto Guida}{address={ICRANet, Piazzale della Repubblica 10, 65122 Pescara, Italy.}, altaddress={Dipartimento di Fisica, Universit\`a di Roma ``La Sapienza'', P.le Aldo Moro 5, 00185 Roma, Italy.}}

\author{Remo Ruffini}{address={ICRANet, Piazzale della Repubblica 10, 65122 Pescara, Italy.}, altaddress={Dipartimento di Fisica, Universit\`a di Roma ``La Sapienza'', P.le Aldo Moro 5, 00185 Roma, Italy.}}

\begin{abstract}
Within the ``fireshell'' model for the Gamma-Ray Bursts (GRBs) we define a ``canonical GRB'' light curve with two sharply different components: the Proper-GRB (P-GRB), emitted when the optically thick fireshell of electron-positron plasma originating the phenomenon reaches transparency, and the afterglow, emitted due to the collision between the remaining optically thin fireshell and the CircumBurst Medium (CBM). We outline our ``canonical GRB'' scenario, with a special emphasis on the discrimination between ``genuine'' and ``fake'' short GRBs.
\end{abstract}

\maketitle

We assume that all Gamma-Ray Bursts (GRBs), including the ``short'' ones, originate from the gravitational collapse to a black hole \citep{2001ApJ...555L.113R,2007AIPC..910...55R,RuffiniTF1}. The $e^\pm$ plasma created in the process of the black hole formation expands as an optically thick and spherically symmetric ``fireshell'' with a constant width in the laboratory frame, i.e. the frame in which the black hole is at rest \citep{1999A&A...350..334R}. We have only two free parameters characterizing the source: the total energy of the $e^\pm$ plasma $E_{e^\pm}^{tot}$ and the $e^\pm$ plasma baryon loading $B\equiv M_Bc^2/E_{e^\pm}^{tot}$, where $M_B$ is the total baryons' mass \citep{2000A&A...359..855R}. These two parameters fully determine the optically thick acceleration phase of the fireshell, which lasts until the transparency condition is reached and the Proper-GRB (P-GRB) is emitted \citep{2001ApJ...555L.113R,2007AIPC..910...55R}.

The afterglow emission then starts due to the collision between the remaining optically thin fireshell and the CircumBurst Medium (CBM) \citep{2001ApJ...555L.113R,2007AIPC..910...55R,2004ApJ...605L...1B,2005ApJ...620L..23B,2005ApJ...633L..13B}. It clearly depends on the parameters describing the effective CBM distribution: its density $n_{cbm}$ and the ratio ${\cal R}\equiv A_{eff}/A_{vis}$ between the effective emitting area of the fireshell $A_{eff}$ and its total visible area $A_{vis}$ \citep{2002ApJ...581L..19R,2004IJMPD..13..843R,2005IJMPD..14...97R,2007A&A...471L..29D}.

\begin{figure}
\includegraphics[width=0.465\hsize,clip]{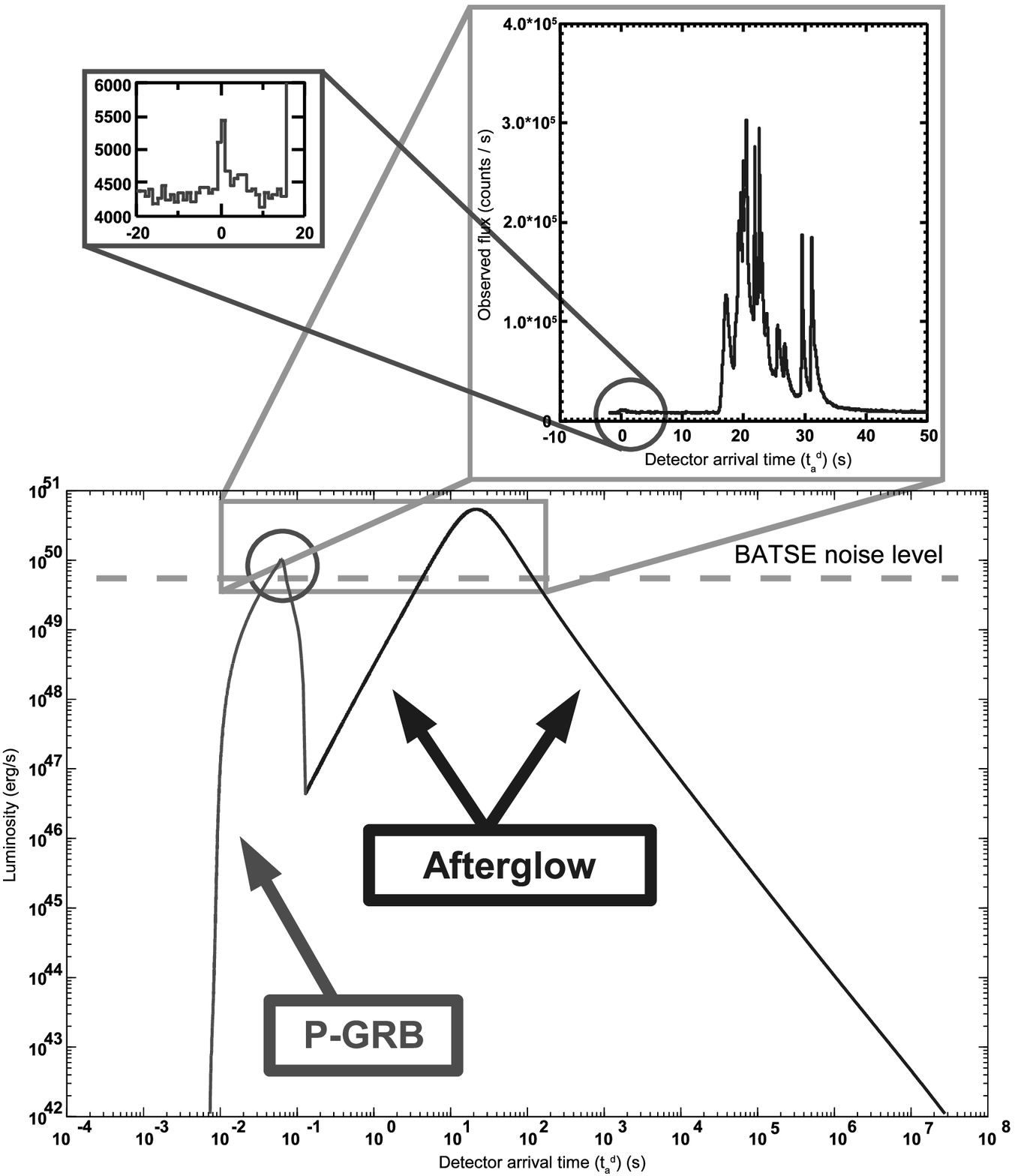}
\includegraphics[width=0.525\hsize,clip]{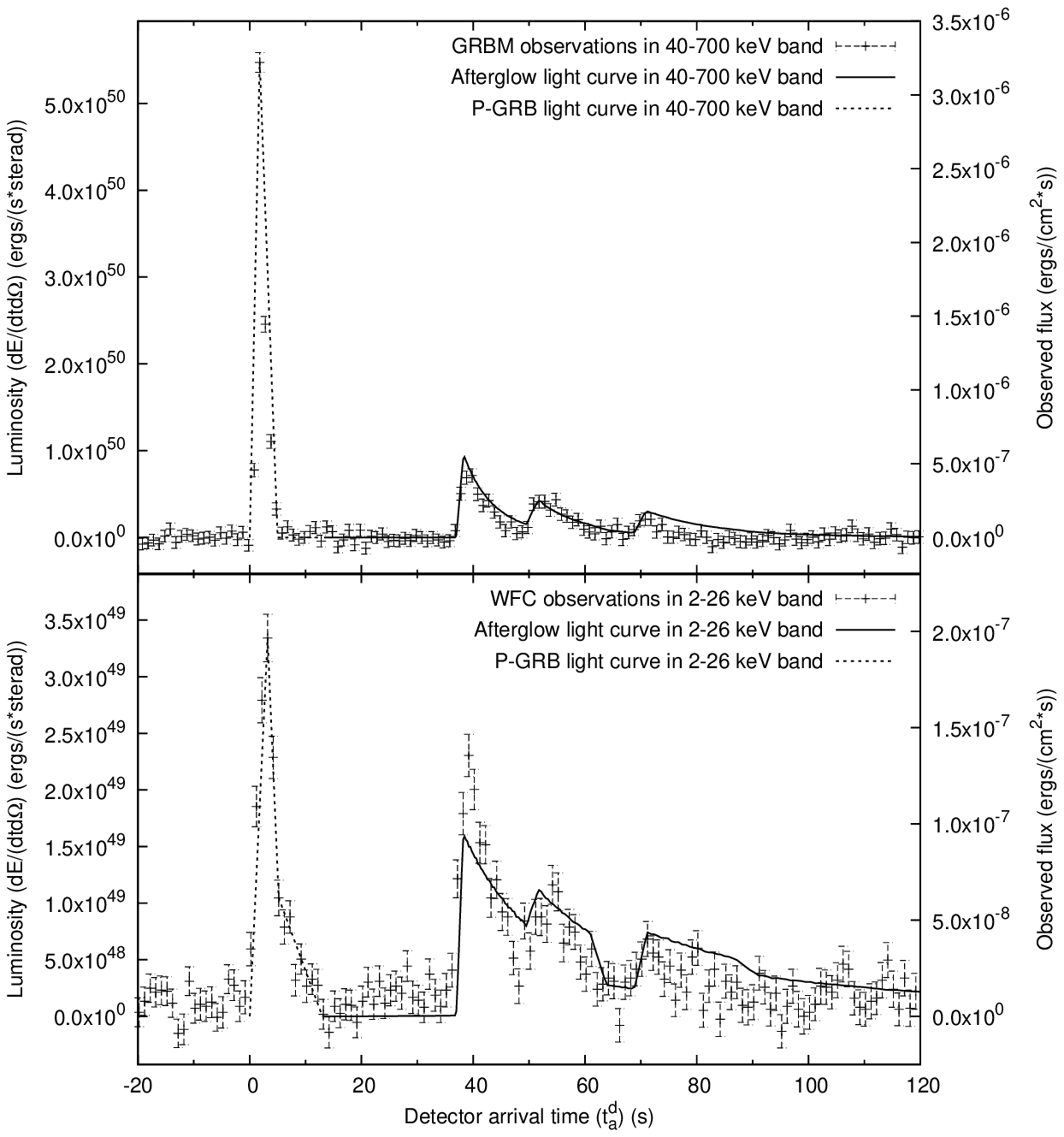}
\caption{{\bf Left:} The ``canonical GRB'' light curve theoretically computed for GRB 991216. The prompt emission observed by BATSE is identified with the peak of the afterglow, while the small precursor is identified with the P-GRB. For this source we have $B\simeq 3.0\times 10^{-3}$ and $\langle n_{cbm} \rangle \sim 1.0$ particles/cm$^3$. Details in \citet{2001ApJ...555L.113R,2002ApJ...581L..19R,2003AIPC..668...16R,2005AIPC..782...42R}. {\bf Right:} The ``canonical GRB'' light curve theoretically computed for the prompt emission of GRB 970228. \emph{Beppo}SAX GRBM ($40$--$700$ keV, above) and WFC ($2$--$26$ keV, below) light curves (data points) are compared with the afterglow peak theoretical ones (solid lines). The onset of the afterglow coincides with the end of the P-GRB (represented qualitatively by the dotted lines). For this source we have $B\simeq 5.0\times 10^{-3}$ and $\langle n_{cbm} \rangle \sim 10^{-3}$ particles/cm$^3$. Details in \citet{2007A&A...474L..13B,grazia_ita-sino}.}
\label{991216_fig}
\end{figure}

Unlike treatments in the current literature \citep[see e.g. Refs.][and references therein]{2005RvMP...76.1143P,2006RPPh...69.2259M}, we define a ``canonical GRB'' light curve with two sharply different components \citep[see Fig.~\ref{991216_fig} and Refs.][]{2001ApJ...555L.113R,2007AIPC..910...55R,2007A&A...474L..13B,bianco_ita-sino}:
\begin{itemize}
\item \textbf{The P-GRB}, which is emitted when the optically thick fireshell becomes transparent and has the imprint of the black hole formation, an harder spectrum and no spectral lag \citep{2001A&A...368..377B,2005IJMPD..14..131R};
\item \textbf{the afterglow}, which presents a clear hard-to-soft spectral evolution in time and which is composed by a rising part, a peak and a decaying tail \citep{2004IJMPD..13..843R,2005ApJ...634L..29B,2006ApJ...645L.109R}; the peak of the afterglow contributes to what is usually called the ``prompt emission'' \citep[see e.g. Refs.][]{2001ApJ...555L.113R,2007A&A...471L..29D,2006ApJ...645L.109R}.
\end{itemize}

\begin{figure}
\includegraphics[width=\hsize,clip]{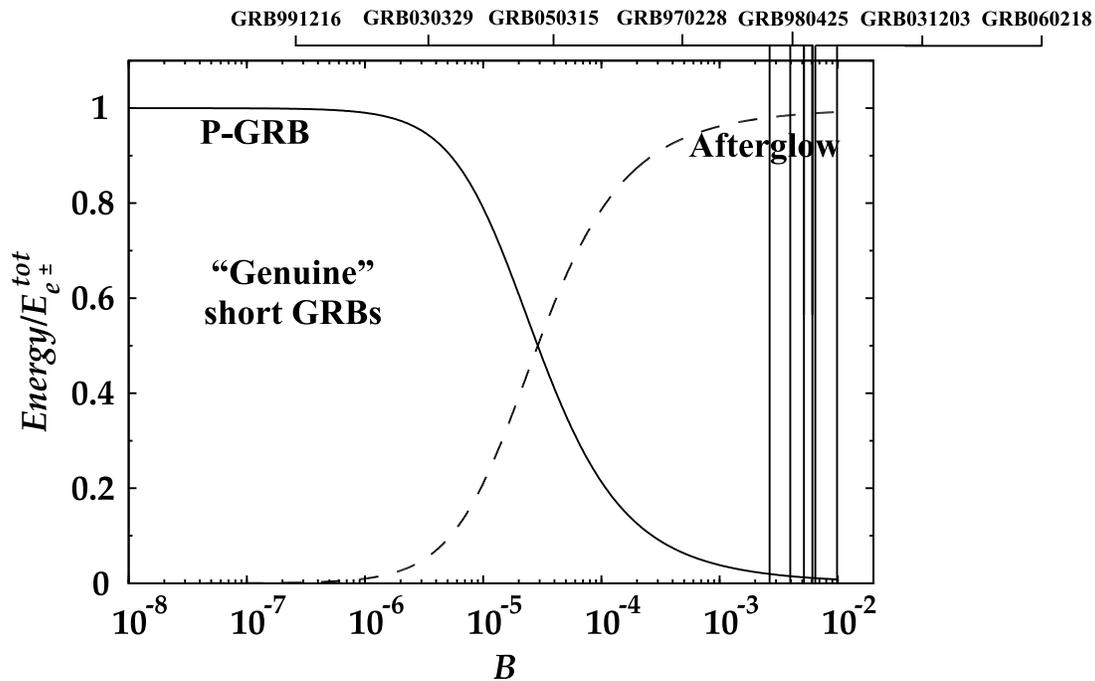}
\caption{The energy radiated in the P-GRB (solid line) and in the afterglow (dashed line), in units of the total energy of the plasma ($E_{e^\pm}^{tot}$), are plotted as functions of the $B$ parameter. Also represented are the values of the $B$ parameter computed for GRB 991216, GRB 030329, GRB 980425, GRB 970228, GRB 050315, GRB 031203, GRB 060218. Remarkably, they are consistently smaller than, or equal to in the special case of GRB 060218, the absolute upper limit $B \lesssim 10^{-2}$ \citep{2000A&A...359..855R}. The ``genuine'' short GRBs have a P-GRB predominant over the afterglow: they occur for $B \lesssim 10^{-5}$ \citep{2001ApJ...555L.113R,2007A&A...474L..13B}.}
\label{figx}
\end{figure}

The ratio between the total time-integrated luminosity of the P-GRB (namely, its total energy) and the corresponding one of the afterglow is the crucial quantity for the identification of GRBs' nature. Such a ratio, as well as the temporal separation between the corresponding peaks, is a function of the $B$ parameter \citep[see Fig.~\ref{figx} and Ref.][]{2001ApJ...555L.113R}. When $B \lesssim 10^{-5}$, the P-GRB is the leading contribution to the emission and the afterglow is negligible: we have a ``genuine'' short GRB \citep{2001ApJ...555L.113R}. When $10^{-4} \lesssim B \lesssim 10^{-2}$, instead, the afterglow contribution is generally predominant \citep[for the existence of the upper limit $B \lesssim 10^{-2}$ see Ref.][]{2000A&A...359..855R}. Still, this last case presents two distinct possibilities: the afterglow peak luminosity can be either \textbf{larger} or \textbf{smaller} than the P-GRB one.

\begin{figure}
\includegraphics[width=\hsize,clip]{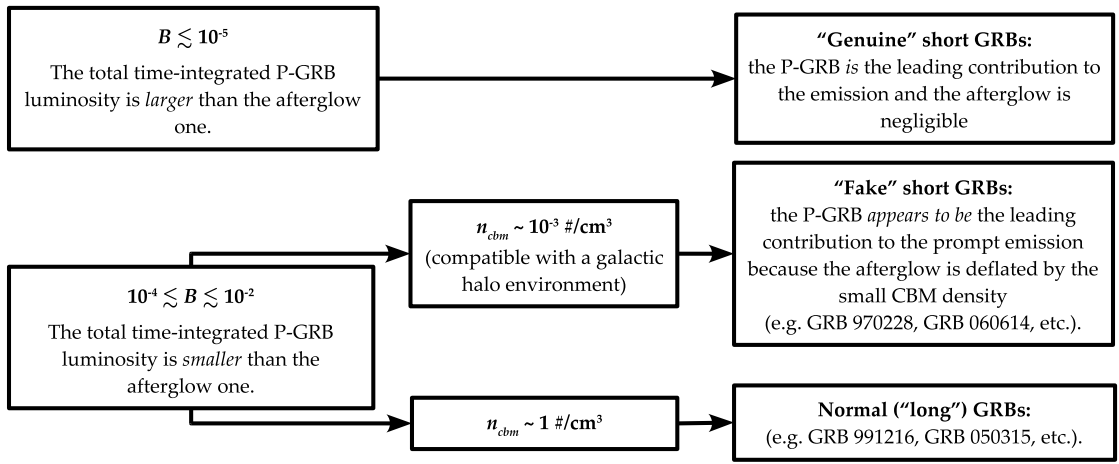}
\caption{A sketch summarizing the ``canonical GRB'' scenario.}
\label{canonical}
\end{figure}

The simultaneous occurrence of an afterglow with total time-integrated luminosity larger than the P-GRB one, but with a smaller peak luminosity, is indeed explainable in terms of a peculiarly small average value of the CBM density ($n_{cbm} \sim 10^{-3}$ particles/cm$^3$), compatible with a galactic halo environment. Such a small average CBM density ``deflates'' the afterglow peak luminosity \citep[see Fig.~\ref{991216_fig} and Refs.][]{2007A&A...474L..13B,grazia_ita-sino}. Of course, such a deflated afterglow lasts longer, since the total time-integrated luminosity in the afterglow is fixed by the value of the $B$ parameter \citep[see above, Fig.~\ref{figx} and Refs.][]{2007A&A...474L..13B,bianco_ita-sino}. Therefore, GRBs belonging to this new class, characterized by a high value of the $B$ parameter and a very small CBM density, inherit their peculiar features from the external environment and not from the intrinsic nature of the source. In this sense, they are only ``fake'' short GRBs \citep{2007A&A...474L..13B,bianco_ita-sino}. Such a class is the one identified by \citet{2006ApJ...643..266N}, includes GRB 060614 and has GRB 970228 as a prototype \citep[see Fig.~\ref{991216_fig} and Refs.][]{2007A&A...474L..13B,bianco_ita-sino}. The whole ``canonical GRB'' scenario is depicted in Fig.~\ref{canonical}.

\end{document}